\pdfoutput=1
\documentclass{webofc}
\usepackage{amsmath,amsfonts,bbm,caption,environ,epstopdf,float,geometry,graphicx}
\usepackage{hyperref,listings,mathrsfs,mdwlist,siunitx,subcaption,wrapfig}
\usepackage[varg]{txfonts}

\newcommand{\centring}{\centering}

\graphicspath{{./Images/}}
\DeclareGraphicsExtensions{.pdf}

\title{Thermal Transitions in Dense Two-Colour QCD}
\author{\firstname{Dale} \lastname{Lawlor}\inst{1}\fnsep\thanks{\email{dalel487@thphys.nuim.ie}, Speaker}
\and \firstname{Simon} \lastname{Hands}\inst{2}\fnsep\thanks{\email{simon.hands@liverpool.ac.uk}}
\and \firstname{Seyong} \lastname{Kim}\inst{3}\fnsep\thanks{\email{skim@sejong.ac.kr}}
\and\firstname{Jon-Ivar} \lastname{Skullerud}\inst{1,4}\fnsep\thanks{\email{jonivar@thphys.nuim.ie}}
}
\institute{
Department of Theoretical Physics; National University of Ireland, Maynooth; Maynooth; Co. Kildare; Ireland
\and Department of Mathematical Sciences; University of Liverpool; Liverpool; L69 3BX; United Kingdom
\and Department of Physics; Sejong University; Seoul 143--147; Republic of Korea
\and Hamilton Institute; National University of Ireland, Maynooth; Maynooth; Co. Kildare; Ireland
}
\abstract{
The infamous sign problem makes it impossible to probe dense (baryon density $\mu_B>0$) QCD at temperatures near or below
the deconfinement threshold. As a workaround, one can explore QCD-like theories such as two-colour QCD
(QC\textsubscript{2}D) which don't suffer from this sign problem but are qualitively similar to real QCD. Previous
studies on smaller lattice volumes have investigated deconfinement and colour superfluid to normal matter transitions.
In this study we look at a larger lattice volume $N_s=24$ in an attempt to disentangle finite volume and finite
temperature effects. We also fit to a larger number of diquark sources to better allow for extrapolation to zero diquark
source.
}

\begin{document}
\maketitle
\section{Introduction}
The objective of lattice QCD is to solve the path integral for an operator $\mathscr{O}[\Phi]$
\begin{equation}
\langle\mathscr{O}\rangle=\frac{1}{Z}\int\mathscr{D}[\Phi]\mathscr{O}[\Phi]e^{-S[\Phi]}
\end{equation}
where $Z$ is the partition function, $\Phi$ is shorthand for all the fields and $S$ the action. $S$ plays the role of a
probability distribution function allowing us to apply Monte Carlo techniques to solve this integral. Gauge
configurations $U$ are produced with probability weight
\begin{equation}
	e^{-S[U]}=\det{M[U]e^{-S_G[U]}}
\end{equation}
where $M[U]$ is the fermion matrix.
For the $\text{SU}(3)$ gauge group with a non-zero baryon chemical potential $\mu_B$, $\det M[U]$  can take on complex
values, yielding a complex probability density. This means that dense QCD systems such as neutron stars and colour
superconducting states are beyond the scope of current lattice simulation techniques. Alternatives such as reweighting,
analytic continuation, the complex Langevin method, effective field theories and studies of other gauge groups are
currently being employed. An overview of the first four approaches can be found in \cite{guenther2021overview} with
this study using the latter approach.

For an $\text{SU}(2)$ gauge group and an even number of flavours the determinant and Pfaffian are both real and
non-negative for $\mu_B>0$, thus allowing Monte Carlo techniques to be used. This two-colour version of QCD has some
interesting properties, such as the role of baryons being played by bosonic diquarks. But qualitatively it is similar
to three-colour QCD, sharing noticeable features such as confinement and chiral symmetry breaking.

%
\begin{figure}
\includegraphics[width=\linewidth]{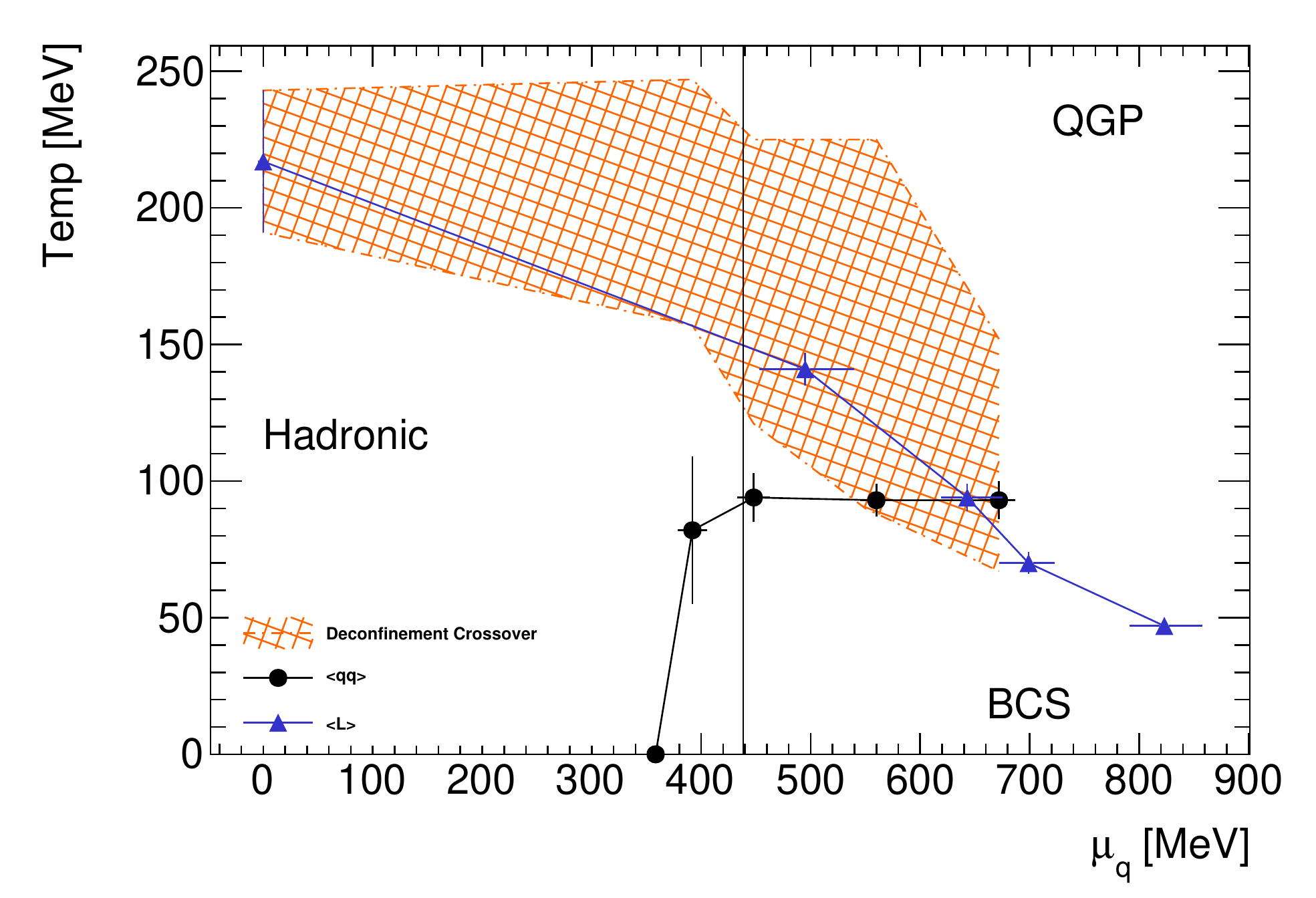}
\captionsetup{width=\linewidth}
\caption{Phase diagram of QC\textsubscript{2}D for $\frac{m_\pi}{m_\rho}=0.80(1)$ \cite{BozEtAl2013}. The orange hatched
area is the deconfinement crossover region, black circles correspond to the  superfluid transition and the blue
triangles the inflexion point of the Polyakov loop from \cite{CotterEtAl2013PhysRev}.}
\label{Phase}
\end{figure}
\section{Simulation details}
For this simulation we are using an unimproved Wilson action with two quark flavours, with simulation parameters from
table \ref{Param Table}. The scale was set by taking the string tension $\sqrt{\sigma}=\SI{440}{\MeV}$ and fitting the
static quark potential to the Cornell form
\begin{equation}\label{Cornell} V(r)=C+\frac{\alpha}{r}+\sigma r\end{equation} at zero chemical potential
\cite{BozEtAl2013,CotterEtAl2013PhysRev}.

We look at a fixed chemical potential $\mu=\SI{443}{\MeV}$ and conduct a temperature sweep. These lattice parameters
were previously used for a chemical potential scan on a spatial extent of $N_s=12$ in \cite{HandsEtAl2010} and both
temperature and chemical potential scans on a spatial extent of $N_s=16$ in \cite{BozEtAl2013EPJA}. The code used for
this simulation was first used in \cite{HandsEtAl2006} and the version used for this run can be found here
\cite{lawlor_dale_2022_7164407} on Zenod on Zenodo.

At low temperatures we expect to find a colour superfluid phase. As the temperature increases we then expect to see a
dense hadronic phase, akin to neutron stars. Beyond that we expect to undergo crossover transition to the deconfined
quark-gluon plasma.

The temperature is given by
\begin{equation}T=\frac{1}{a_\tau N_\tau}\end{equation}
There are two methods of controlling the temperature. By changing the lattice spacing in the temporal extent $a_\tau$ it
is possible to continuously vary the temperature. However this requires setting the scale for every value of $a_\tau$
which is time and resource intensive. Instead we vary the number of sites along the time direction and use a fixed
$a_\tau$. This allows us to complete a temperature scan without setting the scale for each temperature, but at the cost
of only being able to look at a discrete set of temperatures. Temperatures were scanned at $N_\tau=3\text{--}20$,
giving temperatures in the range \SI{55}{\MeV}--\SI{365}{\MeV}. As is standard practice we take the largest time
extent where $N_\tau>N_s$ to be"zero-temperature", thus use the $N_\tau=20$ value of an observable for zero temperature
subtraction.

At non-zero baryon density, the fermion matrix acquires a non-zero density of very small eigenvalues, slowing down the
computation significantly. Introducing a diquark source $j$ lifts these eigenvalues, with "physical" results recovered
by an extrapolation of $j$ to zero as seen in figure \ref{QQvJ}.  Whereas previous studies have used either two or
three diquark sources, this is the first time we have done a full temperature scan with four sources and have a full
temperature scan with $aj=0.010$.
\begin{table}
\centring
\begin{tabular}{|c|c|}
\hline
Parameter&Value\\
\hline\hline
$\beta$&1.9\\ \hline
$\kappa$&0.1680\\ \hline
$a$ (\si{\femto\m})& \SI{0.178(6)}{\femto\m}\\ \hline 
$a$ (\si{\per\GeV})& \SI{0.9}{\per\GeV}\\ \hline 
$a\mu$&0.400\\ \hline
$\mu$ (\si{\MeV})&\SI{443}{\MeV}\\\hline
$a m_\pi$&0.645(8)\\ \hline
$m_\pi$ (\si{\MeV})&\SI{717(25)}{\MeV}\\ \hline
$m_\pi/m_\rho$&0.805(9)\\ \hline
$N_s$&24\\ \hline
\end{tabular}
\caption{Configurations were generated using the above parameters for a coarse lattice as described in
\cite{HandsEtAl2010}.}
\label{Param Table}
\end{table}
\section{Results}
\subsection{Diquark Condensate}
The observables being measured are described in \cite{HandsEtAl2006}, but as a recap we'll be looking at the diquark
condensate
\begin{equation}
\langle qq\rangle=
\frac{\kappa}{2}\langle\bar{\psi_1}C\gamma_5\tau_2\bar{\psi_2^{tr}}-\psi_2^{tr}C\gamma_5\tau_2\psi_1\rangle
\end{equation}
where $C$ is the charge conjugation matrix. The index of the $\psi$ term denotes flavour. This is the order parameter
for the superfluid phase transition (predicted to be a second order transition).

\begin{figure}
\begin{subfigure}{0.5\textwidth}
\includegraphics[width=\linewidth]{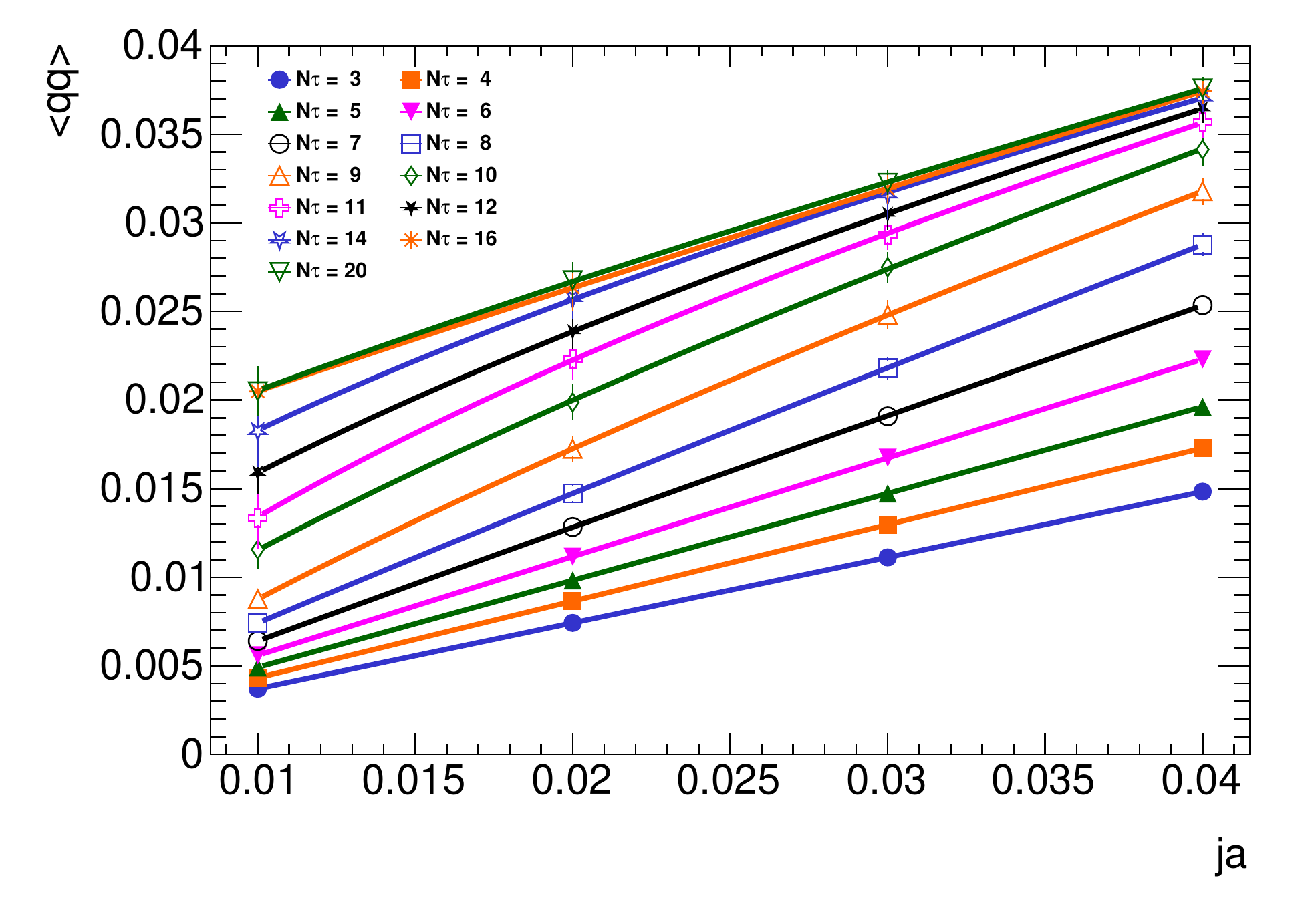}
\captionsetup{width=\linewidth}
\caption{Unrenormalised diquark condensate vs Diquark source.}
\label{QQvJ}
\end{subfigure}
\begin{subfigure}{0.5\textwidth}
\includegraphics[width=\linewidth]{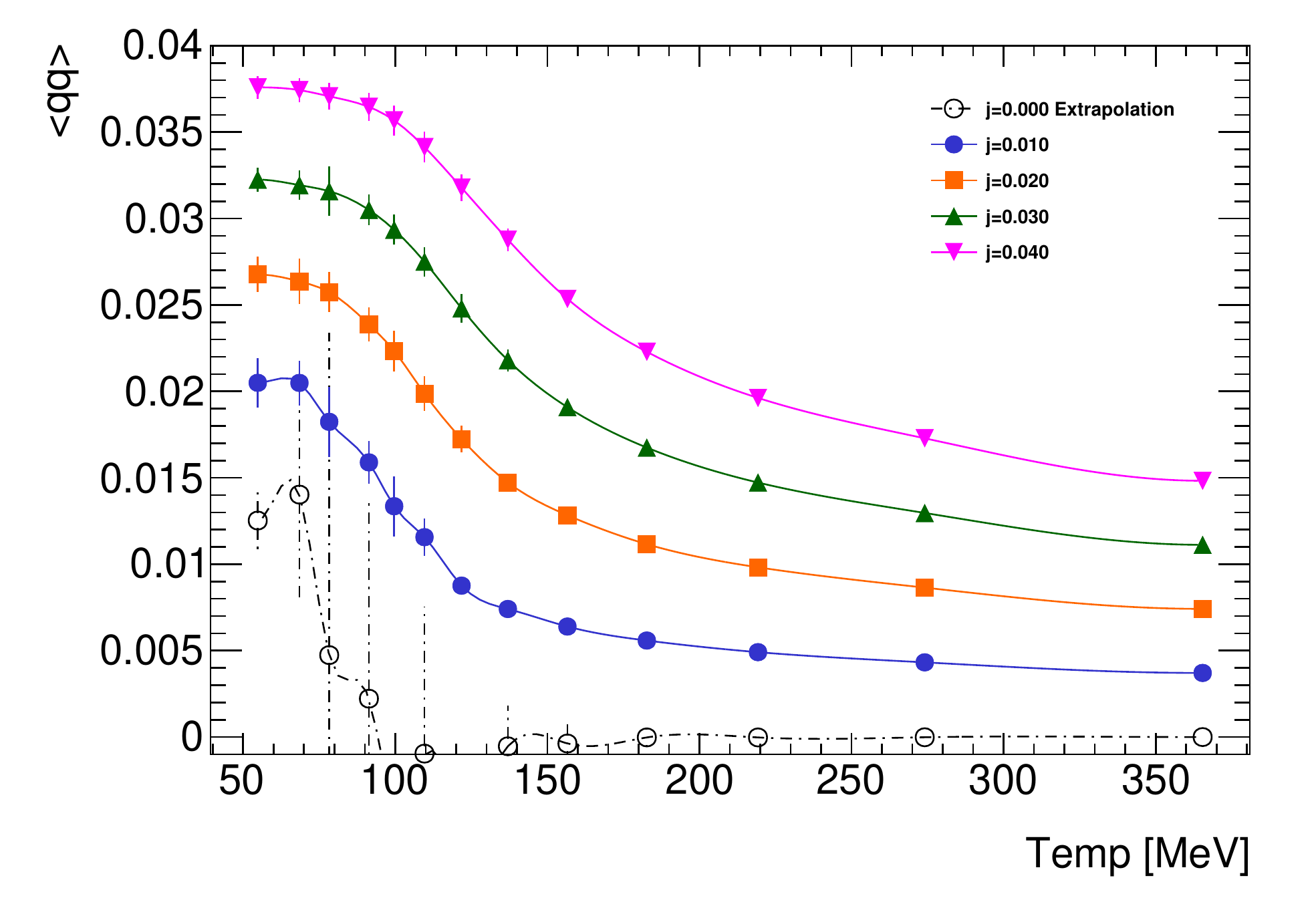}
\captionsetup{width=\linewidth}
\caption{Unrenormalised diquark condensate vs Temperature.}
\label{QQvT}
\end{subfigure}
\caption{Unrenormalised diquark condensate $\langle qq\rangle$ as a function of diquark source $j$ and temperature $T$.
We extrapolated $j$ to zero in figure \ref{QQvJ} to obtain the hollow points in figure \ref{QQvT}.}
\label{QQ}
\end{figure}

Figure \ref{QQvJ} shows the diquark condensate as a function of the diquark source for various temperatures. The fit
$\langle qq\rangle =A+Bj^\alpha$ was used, with the $y$-intercept $A$ being used as the $j=0$ value of the diquark
condensate. This fit has previously been used in \cite{CotterEtAl2013PhysRev,BozEtAl2013EPJA,BozEtAl2020}  on $N_s=12$
and $N_s=16$ lattices using the lattice parameters from this simulation and  with a finer lattice spacing
$a=\SI{0.138(6)}{\femto\metre}$. The $aj=0.01$ values and the larger lattice volume have given us much improved
control of the $j=0$ extrapolation. For low and high temperatures $\langle qq\rangle$ was found to be linear in $j$. But
around the superfluid phase transition the fit is non-linear, with the exponent reaching its minimum around the
transition.

Figure \ref{QQvT} shows the diquark condensate as a function of temperature for various diquark sources. The hollow 
points correspond to the zero diquark extrapolation. We used a cubic spline for the interpolation. A linear fit of the
inflexion points of the three lowest diquark sources suggests the superfluid phase transition occurs around 
$T\sim\SI{84}{\MeV}$, significantly lower than the deconfinement crossover. This indicates that the superfluid phase
transition is indeed distinct from the deconfinement crossover as seen in figure \ref{Phase}. However an improved action
and more data are required to nail down the transition temperature more precisely, in addition to an error analysis to
give proper bounds on the transition temperature.
\subsection{Thermodynamics} 
\begin{figure}
\begin{subfigure}{0.5\textwidth}
\includegraphics[width=\linewidth]{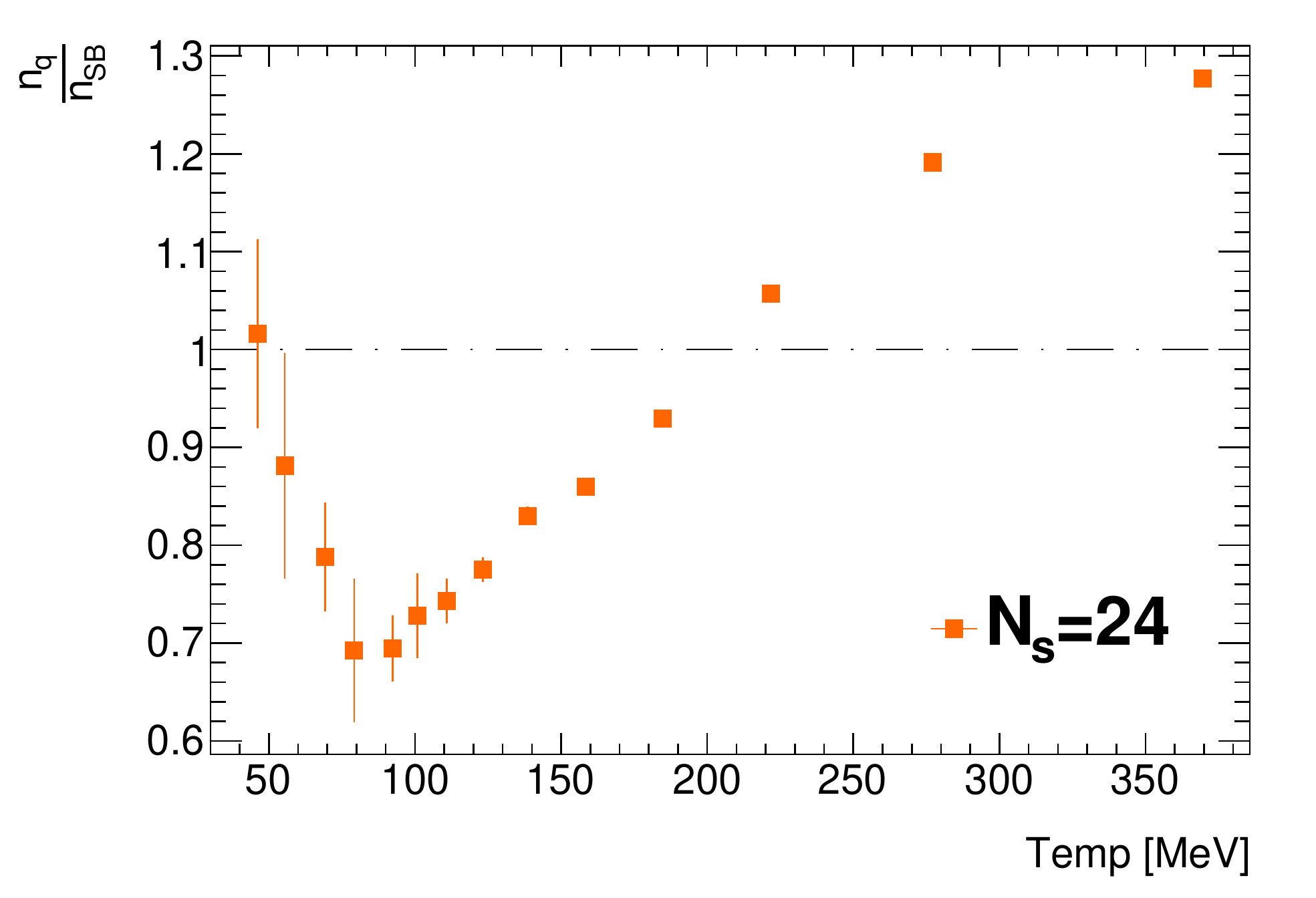}
\captionsetup{width=\linewidth}
\caption{Quark number density normalised by the non-interacting number density.}
\label{n_on_nSB}
\end{subfigure}
\begin{subfigure}{0.5\textwidth}
\includegraphics[width=\linewidth]{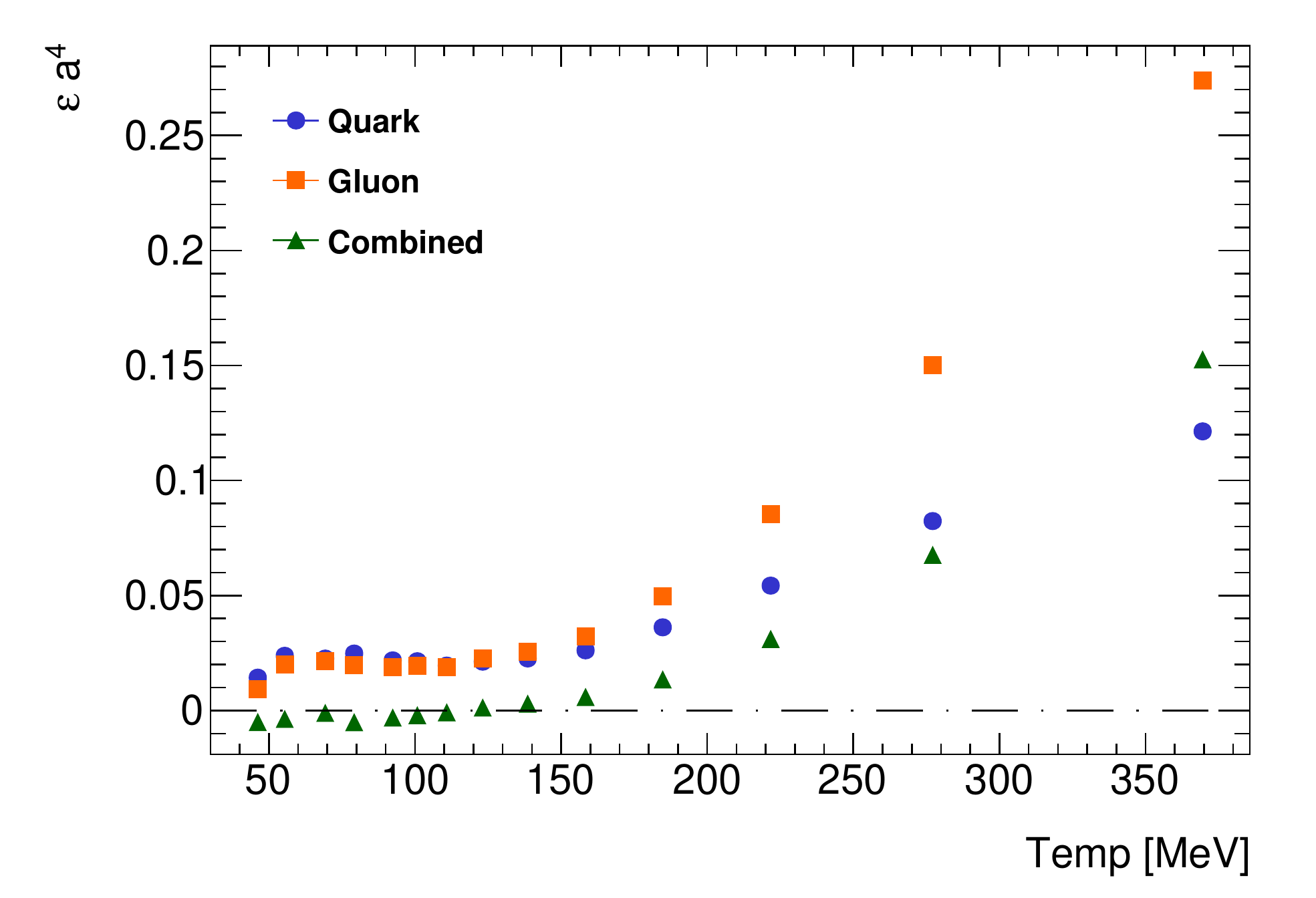}
\captionsetup{width=\linewidth}
\caption{Unrenormalised energy density.}
\label{Energy_Density}
\end{subfigure}
\captionsetup{width=\linewidth}
\caption{Quark number density. These result are compatible with previous results on a smaller volume in
\cite{BozEtAl2013}.}
\label{Num_Densities}
\end{figure}
\begin{figure}
\begin{subfigure}{0.5\textwidth}
\includegraphics[width=\linewidth]{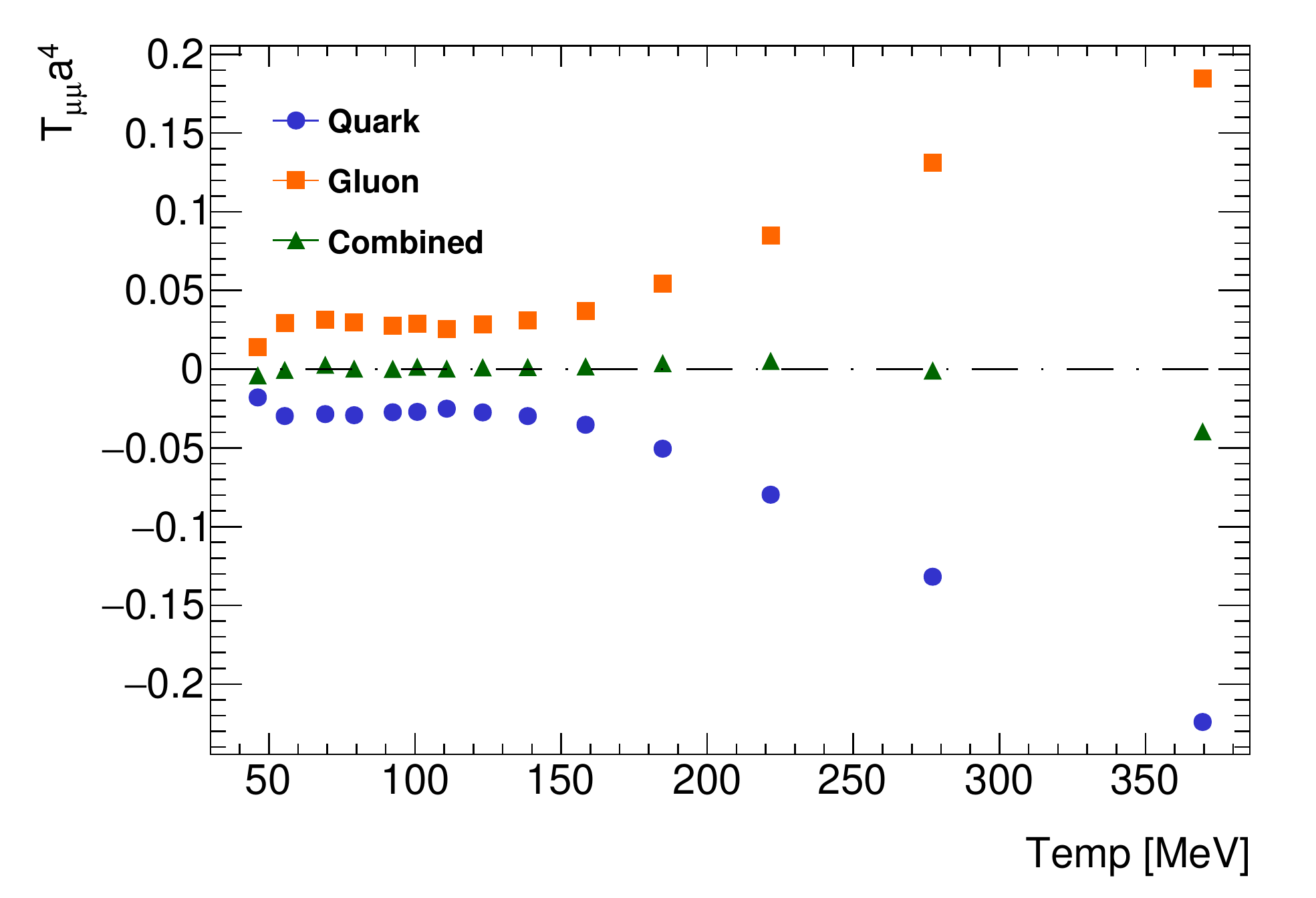}
\captionsetup{width=\linewidth}
\caption{Trace anomaly.}
\label{Trace-Anomaly}
\end{subfigure}
\begin{subfigure}{0.5\textwidth}
\includegraphics[width=\linewidth]{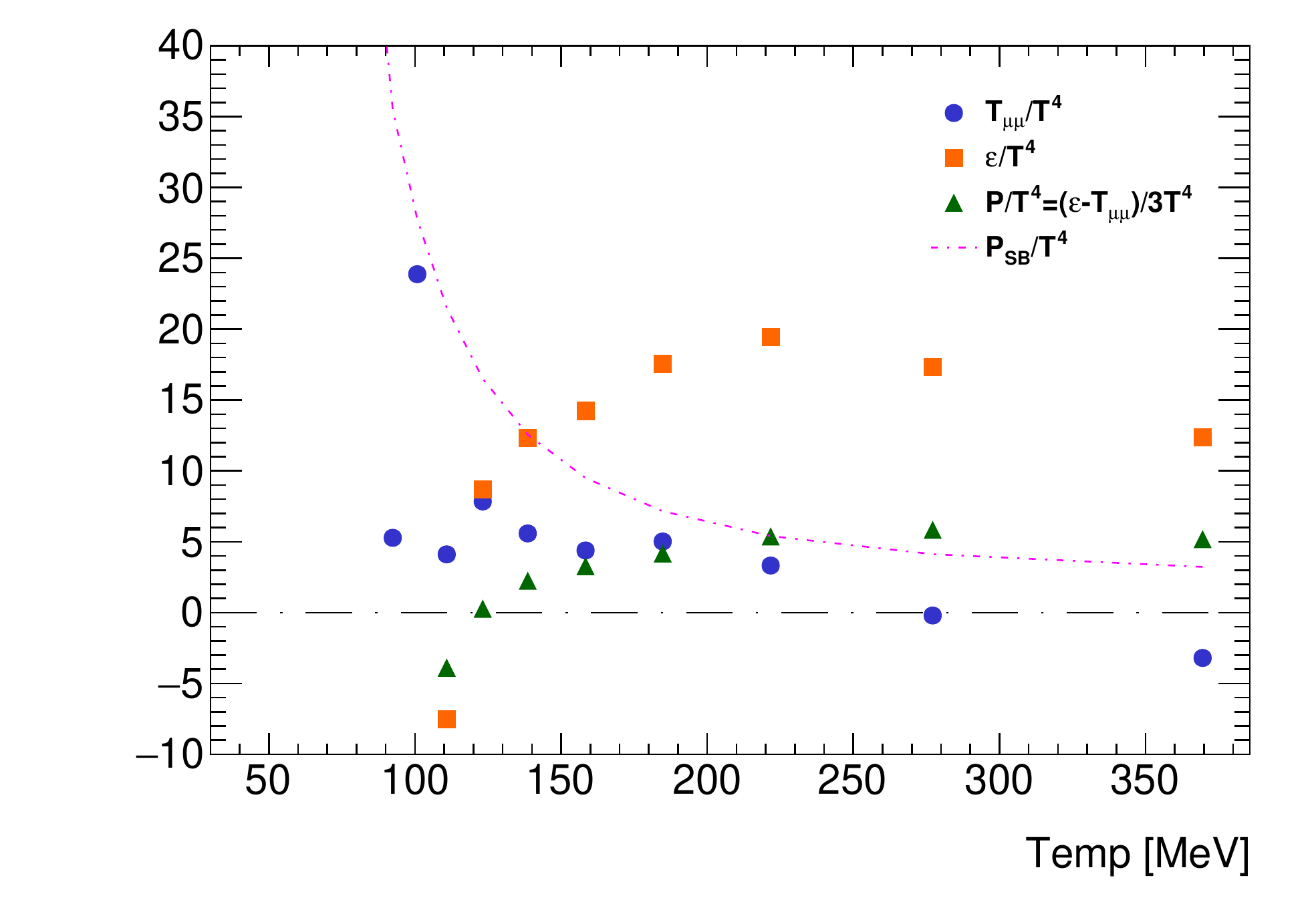}
\captionsetup{width=\linewidth}
\caption{Dimensionless thermodynamic observables.}
\label{Sanity}
\end{subfigure}
\captionsetup{width=\linewidth}
\caption{Thermodynamic observables.}
\label{Trace-Anom-Figure}
\end{figure}
We are also interested in thermodynamic observables such as the quark number density $n_q$, the quark energy
$\varepsilon_q$ and the trace anomaly
\begin{equation}T_{\mu\mu}=\varepsilon-3P\end{equation}
We evaluated the pressure and energy density the using derivative the method, with the Karsch coefficients calculated
in \cite{CotterEtAl2013PhysRev}. These results are qualitatively consistent with recent results from
\cite{IidaEtItou2022velocity,FujimotoEtAl2022TraceAnomaly}. The energy density exhibits similar behaviour. The dip in
the density in figure \ref{n_on_nSB} is in the same region as the superfluid phase transition. This is remarkable as
$\frac{n_q}{n_{SB}}$ is not an order parameter for the phase transition.

The trace anomaly itself is smaller than expected. Also interesting is how the quark and gluon contributions nearly
cancel each other out at most temperatures. This may be due to the choice of regularisation. The trace anomaly is
constant and consistently zero up until $T\sim\SI{150}{\MeV}$, suggesting that we are in a conformal r\'egime. Is this
a sign of a second transition from conformal to non-conformal physics?
\subsection{Chiral and Deconfinement Transition}
\begin{figure}
\begin{subfigure}{0.5\textwidth}
\includegraphics[width=\linewidth]{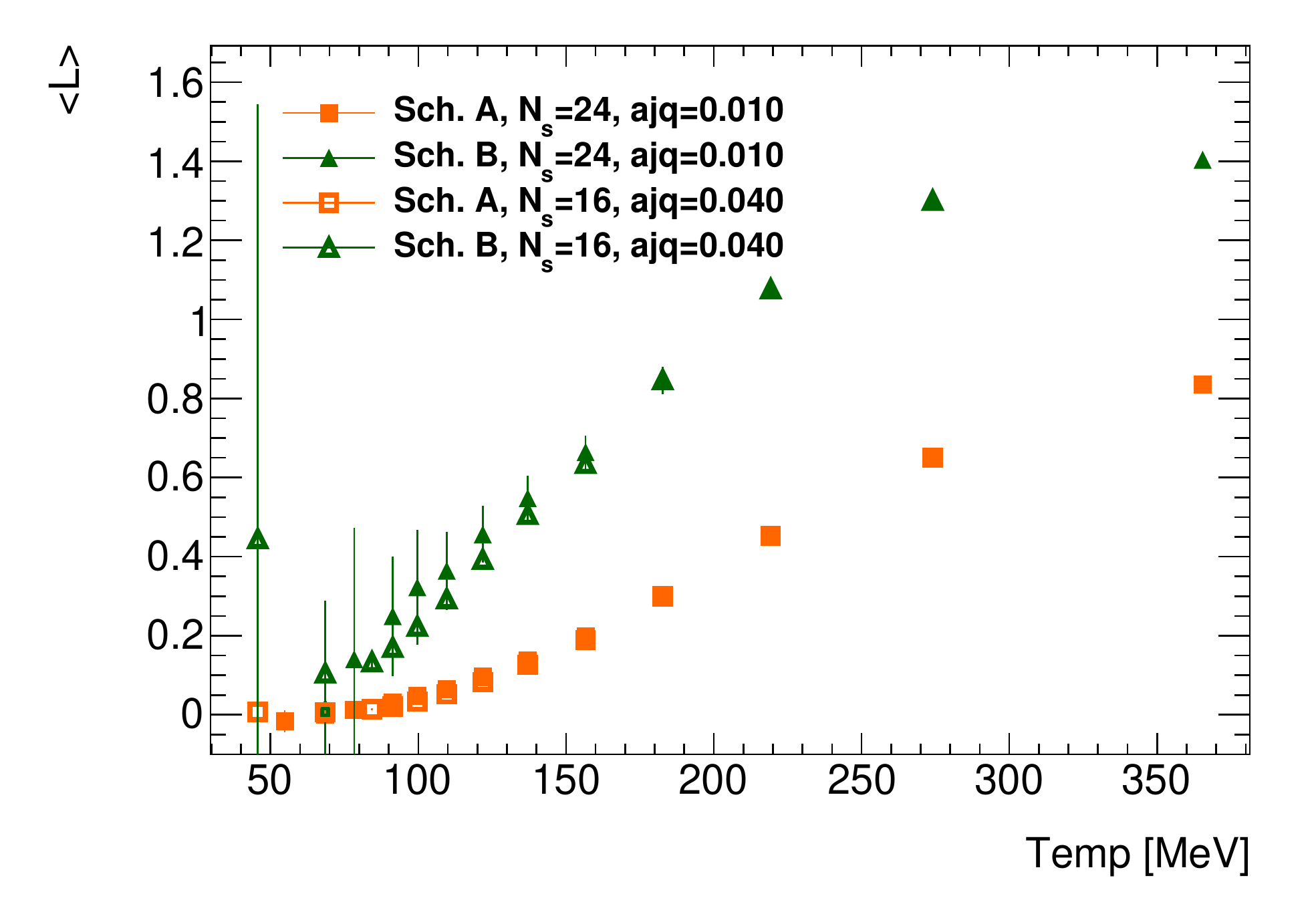}
\captionsetup{width=\linewidth}
\caption{Renormalised Polyakov loop}\label{Poly}
\end{subfigure}
\begin{subfigure}{0.5\textwidth}
\includegraphics[width=\linewidth]{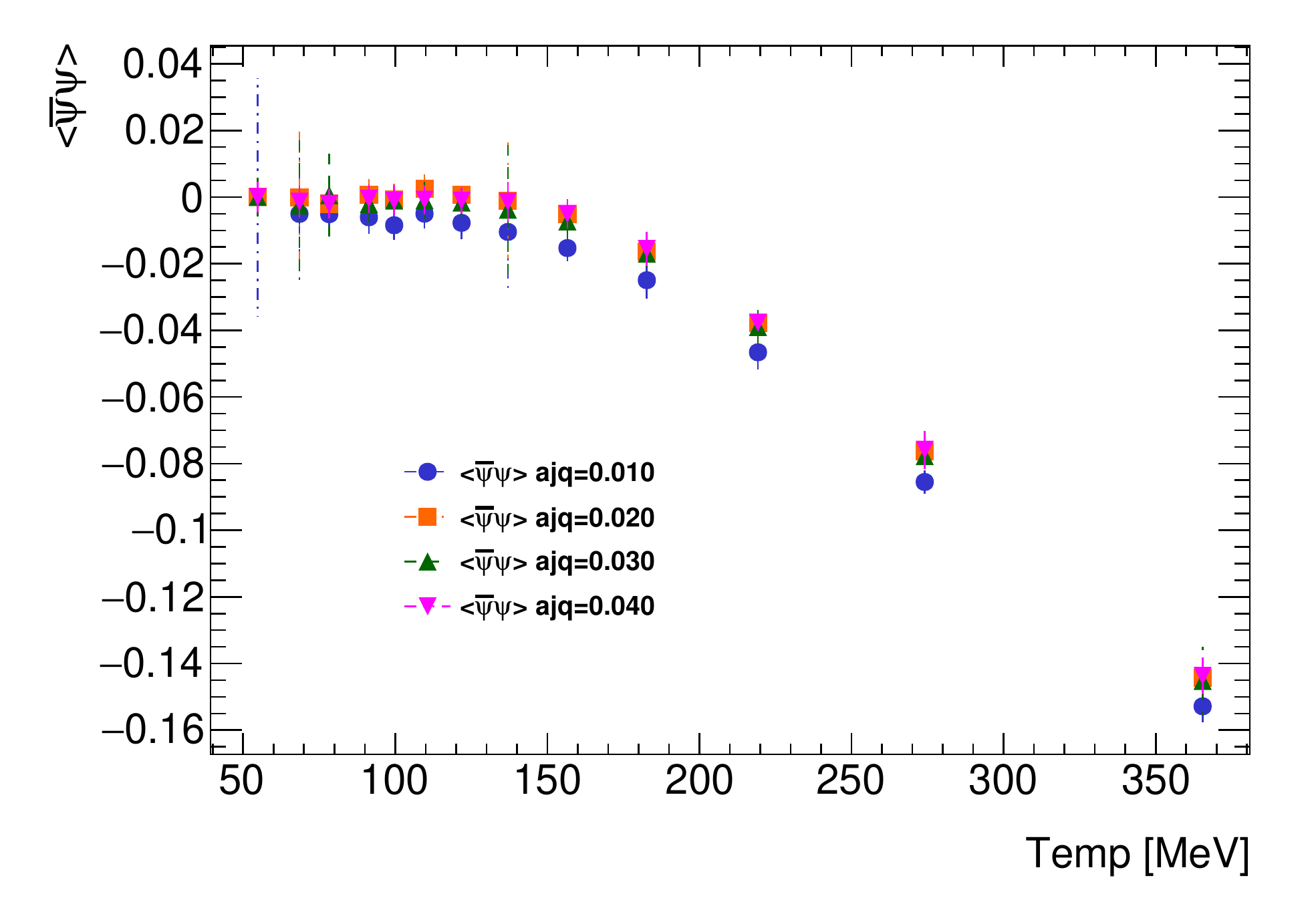}
\captionsetup{width=\linewidth}
\caption{Subtracted but unrenormalised chiral condensate}
\label{Chiral-Condensate}
\end{subfigure}
\captionsetup{width=\linewidth}
\caption{Bosonic observables}
\end{figure}
Another quantity of interest is the Polyakov loop expectation value, which is the order parameter for the deconfinement
crossover in pure gauge theory. The Polyakov loop was renormalised using the procedure described in
\cite{Bors_nyi_2012_PLB} with the renormalisation constants previously calculated on a smaller volume in
\cite{CotterEtAl2013PhysRev}. The Polyakov loop renormalisation is temperature dependent, unlike that of the diquark
condensate.

\begin{equation} L_R\left(T,\mu\right)=Z_L^{N_\tau} L_0\left(\frac{1}{a_\tau N_\tau},\mu\right)\end{equation}
We consider two renormalisation schemes to determine $Z_L$
\begin{table}[H]
\centring
\begin{tabular}{c c c}
\textbf{Scheme A:} & $L_R(N_\tau=4,\mu=0)=0.5$ & $\Rightarrow Z_L=1.37495$\\
\textbf{Scheme B:} & $L_R(N_\tau=4,\mu=0)=1.0$ & $\Rightarrow Z_L=1.15619$
\end{tabular}
\end{table}
The chiral condensate with Wilson fermions has both multiplicative and additive renormalisations. The additive and
multiplicative renormalisations are a constant shift and constant factor respectively. Figure \ref{Chiral-Condensate}
has undergone additive renormalisation by subtracting the zero temperature value of $\langle qq \rangle$, but
multiplicative renormalisation has not been carried out. This can be done using the proceedure found in
\cite{aarts2020properties,Giusti_1999,Bors_nyi_2012_JHEP} but would merely amount to an overall rescaling of the data
in figure \ref{Chiral-Condensate}. The change in behaviour from constant to decreasing at  $T\sim\SI{150}{\MeV}$
suggests that the crossover coincides with the deconfinement crossover, not the superfluid transition.

\section{Outlook}
The larger lattice volume has shown us that finite volume errors are small for these parameters. 
The superfluid phase transition and deconfinement crossover are distinct. Further investigation is needed into the
unexpectedly small trace anomaly. Work is underway to implement a Symanzik improved fermion action. We have also started
tuning for a finer lattice with lighter quarks. Combined with the improved action these "light-fine" quarks will give us
better control over errors in addition to providing more physically realistic measurements.
\section*{Acknowledgements}
\begin{acknowledgement}
D. Lawlor would like to acknowledge support from the National University of Ireland, Maynooth's John and Pat Hume
Scholarship.

The authors wish to acknowledge the Irish Centre for High-End Computing (ICHEC) for the provision of computational
facilities and support.

This work was performed using the DiRAC Data Intensive service at Leicester, operated by the University of Leicester
IT Services, which forms part of the STFC DiRAC HPC Facility (\url{www.dirac.ac.uk}). The equipment was funded by BEIS
capital funding via STFC capital grant ST/K000373/1 and ST/R002363/1 and STFC DiRAC Operations grant ST/R001014/1.
DiRAC is part of the National e-Infrastructure.
\end{acknowledgement}
\bibliography{./BibTeX.bib}
\end{document}